\documentclass[a4paper,fleqn,usenatbib]{mnras}

\usepackage{mathptmx}

\usepackage[T1]{fontenc}
\usepackage{ae,aecompl}

\usepackage{graphicx}	
\usepackage{amsmath}	
\usepackage{amssymb}	
\usepackage[retainorgcmds]{IEEEtrantools}
\usepackage{placeins}

\usepackage{afterpage}  
\usepackage[usenames, dvipsnames]{color}
\definecolor{ao(english)}{rgb}{0.93, 0.53, 0.18}
\definecolor{green}{rgb}{0.0, 0.8, 0.6}

\usepackage[normalem]{ulem}  
\def\be{\begin{equation}}
\def\ee{\end{equation}}
\def\bi{\begin{itemize}}
\def\ei{\end{itemize}}

\title[Effects of chaos on the detectability of stellar streams]{Effects of chaos on the detectability of stellar streams}

\author[Mestre et al.]{
Mart{\'i}n Mestre~$^{1,2}$\thanks{E-mail: mmestre@fcaglp.unlp.edu.ar},
Claudio Llinares~$^{3,4}$,
Daniel D. Carpintero~$^{1,2}$
\\
$^{1}$Instituto de Astrof{\'i}sica de La Plata (CONICET-UNLP), Argentina \\
$^{2}$Facultad de Ciencias Astron{\'o}micas y Geof{\'i}sicas de La Plata (UNLP), Argentina \\
$^{3}$Institute for Computational Cosmology, Department of Physics, Durham University, Durham DH1 3LE, U.K. \\
$^{4}$Institute of Cosmology and Gravitation, University of Portsmouth, Dennis Sciama Building, Portsmouth PO1 3FX, United Kingdom
}

\date{Accepted XXX. Received YYY; in original form ZZZ}

\pubyear{2019}

\begin{document}
\label{firstpage}
\pagerange{\pageref{firstpage}--\pageref{lastpage}}
\maketitle

\begin{abstract}  
  Observations show that stellar streams originating in satellite dwarf galaxies are frequent in the Universe.  While such events are predicted by theory, it is not clear how many of the streams that are generated are washed out afterwards to the point in which it is imposible to detect them.  Here we study how these diffusion times are affected by the fact that typical gravitational potentials of the host galaxies can sustain chaotic orbits.  We do this by comparing the behaviour of simulated stellar streams that reside in chaotic or non-chaotic regions of the phase-space.  We find that chaos does reduce the time interval in which streams can be detected.  By analyzing detectability criteria in configuration and velocity space, we find that the impact of these results on the observations depends on the quality of both the data and the underlying stellar halo model.
 For all the stellar streams, we obtain a similar upper limit to the detectable mass.
\end{abstract}




\section{Introduction}
Observations show that the destruction of dwarf satellite galaxies due to tidal fields associated to the gravitational potential of the host galaxy is frequent \citep{2018MNRAS.475.1549M,2009MNRAS.394.1956C,1993MNRAS.261..921K}.
Stellar streams are unavoidable consequences, either permanent or transitory, of this type of interaction~\citep{2018A&A...614A.143M,2018ApJ...862..114S,2008ApJ...689..184M}. 
Although the main mechanism of stream formation and its relevance to the understanding of the origin and evolution of the Milky Way (MW) are well understood \citep[e.g.][]{1977MNRAS.180...71D,1972MNRAS.157..309W, 1999Natur.402...53H,1999MNRAS.307..495H},  there is still much to be done on the application of stellar streams to galactic archaeology and cosmology.  For instance, the properties of streams can be used to put constraints on the structure and evolution of the stellar and dark matter haloes of galaxies \citep{2017A&A...598A..58H,2018MNRAS.481.3442M,2016PhRvL.116l1301B} as well as on the nature of the dark matter particles themselves~\citep{2018JCAP...07..061B}. Regarding the intrinsic aspects of stellar streams, \citet{2015MNRAS.450..575A} made a theoretical and numerical  characterization of three observable quantities: the speed of the stream's growth,  the internal coherence of the stream, and its thickness or opening angle within and outside the orbital plane. Moreover,~\citet{2015MNRAS.454.2472H} and~\citet{2019MNRAS.487..318K} have demonstrated that it is posible to relate orbital parameters of the orbits with their morphology: shell--like or string--like shapes.  

The study of stellar streams can be simplified by taking into account that the dynamics of the stars once they have been pulled out of the satellite depends almost exclusivelly on the properties of the gravitational potential of the host (i.e. self--gravity of the streams is negligible).  Thus, we can apply tools and theorems that were built in the past for Hamiltonian systems with few degrees of freedom.  Among them, chaos theory~\citep{2002ocda.book.....C} is a well developed branch of study  that has played a big role in many astrophysical theories, such as in Solar System  dynamics~\citep{2010CRPhy..11..651M,Tsiganis2005}, multi--planet extrasolar systems~\citep{2013MNRAS.433..928M,2012RAA....12.1044B}, barred spiral galaxies~\citep{2012CeMDA.113...81C,2006MNRAS.373..280V} and galaxy haloes~\citep{2017MNRAS.466.3876Z,2016A&A...595A..68V,2014MNRAS.438.2871C,2007LNP...729..297E,2000MNRAS.319...43S,1998ApJ...506..686V,1996ApJ...471...82M}.
And the field of stellar streams belongs to this group since the early works
of~\citet{doi:10.1196/annals.1350.011},~\citet{2001A&A...373..511F} and~\citet{2000AJ....119..800D}. 
Recently, \citet{2018MNRAS.478.4052M,2015MNRAS.453.2830M} have shown that
for stellar streams in the solar neighbourhood the effect of the chaotic
mixing is negligible, in the sense that the proportion of erased substructure
during a Hubble time is not significant.
This is not in contradiction with the evidence about chaos affecting
the morphology of streams whose progenitors are globular
clusters~\citep{2019arXiv191000592B,2016MNRAS.455.1079P,2016ApJ...824..104P,Sesar_2015,2015MNRAS.452..301F,2015ApJ...799...28P}.
This type of tidal streams are thinner and dynamically colder~\citep{2019ApJ...883...87P,2013MNRAS.436.2386L} than those originated from dwarf satellite galaxies
and the overall impact of chaos on streams born from these heavier progenitors is  still uncertain.

Here we study the relation between chaos and streams originated in dwarf--sized satellite galaxies embedded in the dark halo of a host galaxy similar to the MW. We do this by comparing the evolution of simulated streams that reside in chaotic and non-chaotic (regular) regions of the potential of the host.  We measure the detectability of the streams by applying a simple criterion in either configuration or velocity space.
We find that the presence of chaos does have an impact on the stream evolution. Streams that evolve in chaotic regions of the potential have shorter spans of detectability.  We also find that the significance of this effect depends on details of the detectability criterion and
the quality of the underlying stellar halo model.

A detailed description of the set up of the simulations is provided in Section~\ref{section:simulations}.  Our analysis techniques and definition of the detectability criterion that we applied to the data are presented in Section~\ref{section:analysis}. The results of our work are given in Section~\ref{section:results}. Finally we discuss the results and conclude  in Section \ref{section:conclusions}.

\section{Simulations}
\label{section:simulations}
\begin{figure*}
\centering
\includegraphics[width=\textwidth]{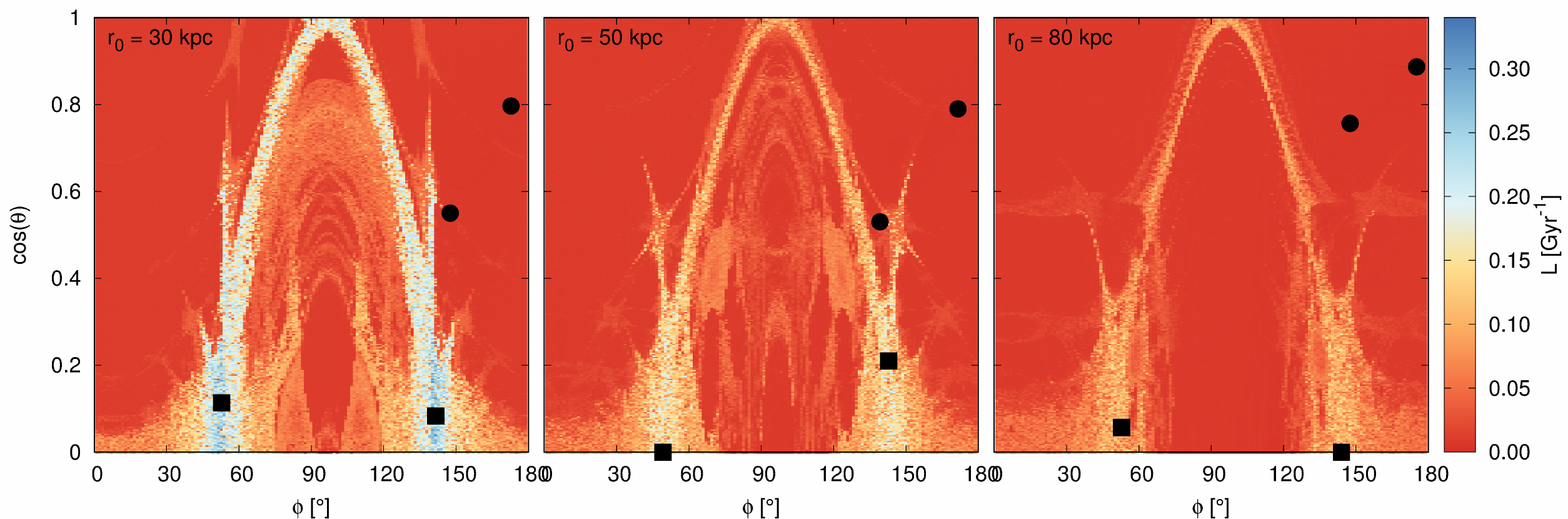}
\caption{Chaoticity and selection of initial conditions for the satellites in configuration space.  Each panel corresponds to a different value of the galactocentric distance $r_0$. The background colours correspond to the Lyapunov exponent, $L$, used to classify the orbits as regular (red) or chaotic (orange, blue and white).  Each black point corresponds to the initial position
  of the centre of mass of a progenitor (circles and squares for regular and chaotic orbits respectively).
  In all the cases, the inital velocity of the center of mass of the satellites is a horizontal vector pointing to the right.} 
    \label{fig:LI_progenitors_rApo-all}
\end{figure*}

We study the impact that chaos has in the evolution of tidal streams by analysing simulated streams.  The simulations consist in the $N$-body integrations of the trajectories of dark matter particles of a dwarf galaxy in a smooth gravitational potential that represents the Milky Way.  The integrations were made with the code \textsc{gadget-2}~\citep{2005MNRAS.364.1105S}.  In order to work with a realistic satellite, we determine the initial conditions of its constituent particles to be consistent with a classical MW satellite~\citep{2010MNRAS.406.1220W} such as Sculptor.

\subsection{Host galaxy}
The host galaxy is defined as a smooth analytic potential which we chose to be a generalized NFW potential~\citep{1997ApJ...490..493N, 2001MNRAS.321..155L}:
\be
\label{nfw_pot}
\Phi(\textbf{x}) =\Phi_{\textsc{nfw}}(k(\textbf{x})) =- \frac{GM_{200}}{f(c)} \frac{\ln(1+k(\textbf{x})/r_s)}{k(\textbf{x})}, 
\ee
where $G$ is the gravitational constant, $M_{200}$ is the virial mass, $c$ is the concentration parameter, $r_s$ is the virial radius ($r_{200}$) in units of the concentration, $k(\textbf{x})$ parametrizes the asphericity of the halo and the free function $f$ is given by $f(c) = \ln(1+c) - c(1+c)^{-1}$.  We ussumed the following parameters:  $G=4.3009\times 10^{-6}$~kpc km$^2$ s$^{-2}$, $M_{200}=1.5\times 10^{12}$~M$_{\odot}$ ($r_{200}=242.40$ kpc), which are consistent with~\citet{2006MNRAS.369.1688D} and~\citet{doi:10.1111/j.1365-2966.2010.16708.x} together with $c=11.27$ in agreement with~\citet{1999dmap.conf..375B} so that $r_s=r_{200}/c=21.51~\rm{kpc}$.

The generalization of the spherical potential is given by a smooth transition between an axisymmetric potential in the inner region and a triaxial one in the outskirts, which is parametrized as follows
\be
k(\textbf{x}) = \left( \frac{1+r_t(\textbf{x})/r_{\rm{g}}}{1+r_a(\textbf{x})/r_{\rm{g}}} \right) r_a(\textbf{x}), \nonumber
\ee
where we assumed a transition radius $r_{\rm{g}}=1.2 r_s \approx 25.81~\rm{kpc}$, as in~\citet{2007MNRAS.377...50H}.   The same parametrization was applied by~\cite{2013ApJ...773L...4V} to a logarithmic potential~\citep[see also][]{2008MNRAS.385..236V}.  The function $r_a$ that parametrizes the axisymmetric dependence was taken from~\cite{2005ApJ...619..800J}:
\be
r_a  = \sqrt{x^2+y^2+(z/q_z)^2}, \nonumber
\ee
where $q_z= 0.93$ is consistent with fits to M-giants selected from the Two Micron All Sky Survey (2MASS) that correspond to tidal debris in the Milky Way, performed by the same authors.

The other free function $r_t$ provides the triaxial dependence and was taken from~\cite{2010ApJ...714..229L}:
\be
\label{lm10a}
r_t  = \sqrt{C_1 x^2+ C_2 y^2+ C_3 xy +(z/q_3)^2}, 
\ee
were
\begin{align}
C_1 & = \frac{\cos^2\alpha}{q_1^2}+\frac{\sin^2\alpha}{q_2^2}, \nonumber \\
C_2 & = \frac{\cos^2\alpha}{q_2^2}+\frac{\sin^2\alpha}{q_1^2}, \nonumber   \\ 
C_3 & = 2\cos\alpha \sin\alpha\left(\frac{1}{q_1^2}-\frac{1}{q_2^2}\right).\nonumber 
\end{align}
For the free parameters in these expresions, we chose values that were obtained by the same
authors as best fit parameters to data
of the Sagittarius dwarf galaxy: $\alpha  = 97^{\circ}$, $q_1 = 1.38$, $q_2 = 1$ and $q_3  = 1.36$.
Although \cite{2013MNRAS.428..912D} found
a model with similar shape and extreme orientation relative to the galactic
disc, it is unstable according to~\cite{2013MNRAS.434.2971D}
and thus improbable in standard cosmological simulations.
The works of~\cite{2015ApJ...799...28P},~\cite{2014MNRAS.439.2678D}
and~\cite{2013ApJ...773L...4V} provide further details about
the unrealistic character of this model (i.e. the~\citealt{2010ApJ...714..229L}
triaxial potential with constant axial ratios).
In spite of this, our aspherical MW halo model given in Eq.~\ref{nfw_pot}
is an example of a non--integrable gravitational potential whose
phase--space is inhabited by regular and chaotic orbits, thus
serving for the purpose of this research.

\subsection{Orbits of the progenitors}
\label{sec:orbits}

\begin{table*}
  \caption{Initial conditions for the centre of mass of the progenitors, radius of the equivalent circular orbit, largest Lyapunov characteristic exponent, classification and Lyapunov time (only for the chaotic ones), minimum pericentric distance, mean pericentric distance and its standard deviation.}
  \label{table_ics}  
  \begin{tabular}{ | r | c | c | c | c | c | c | c | c | c | c | c |} 
    \hline  $\#$ Sat & $r_0$ [kpc] & $\phi_0$ [$^{\circ}$] & $\cos \theta_0$  & ${\rm{v}}^{\phi}_0$ [km s$^{-1}$]& $r_c$ [kpc]  & $L$ [Gyr$^{-1}$] & Classification & $T_{\rm{L}}$ [Gyr] &
    $r_{\mathrm{min}}$ & $\langle r_{\mathrm{p}} \rangle$ & $\sigma_{\mathrm{p}}$ \\ 
    \hline
     1& 30 &  52.8 & 0.113 & 67.46 & 19.87 & 0.3344 & Chaotic & 2.99 & 2.13 & 9.31 & 3.12 \\ 
     2& 30 & 141.6 & 0.083 & 67.46 & 19.87 & 0.3229 & Chaotic & 3.12 &  1.97 &  9.04&   3.29\\ 
     3& 30 & 172.8 & 0.797 & 67.46 & 19.87 & 0.0053 & Regular & --- & 7.06&  11.25 & 1.86\\ 
     4& 30 & 147.6 & 0.550 & 67.46 & 19.87 & 0.0060 & Regular & --- & 5.62 &  9.91 &  2.09\\
    \hline
     5& 50 &  49.2 & 0.000 & 70.06 & 32.50 & 0.2380 & Chaotic & 4.16 & 4.61&  12.37 &  5.37 \\ 
     6& 50 & 142.8 & 0.210 & 70.06 & 32.50 & 0.2271 & Chaotic & 4.42 & 4.58 &  12.79 &  3.43 \\ 
     7& 50 & 139.2 & 0.530 & 70.06 & 32.50 & 0.0047 & Regular & --- & 5.74 &  13.07 &  3.50 \\ 
     8& 50 & 171.6 & 0.790 & 70.06 & 32.50 & 0.0054 & Regular & --- & 12.14 & 16.33 &  1.86  \\
    \hline
     9& 80 &  52.8 & 0.057 & 69.57 & 51.05 & 0.1766 & Chaotic & 5.72 & 7.80 &  17.79 &  6.17\\ 
     10& 80 & 144.0 & 0.000 & 69.57 & 51.05 & 0.1711 & Chaotic & 5.85 &3.54 &  12.83 &  7.34\\ 
     11& 80 & 147.6 & 0.757 & 69.57 & 51.05 & 0.0041 & Regular & --- &16.48 &  20.29 &  3.36\\ 
     12& 80 & 175.2 & 0.887 & 69.57 & 51.05 & 0.0044 & Regular &  --- & 18.644 &  21.25 &  2.50\\
     \hline
  \end{tabular}
\end{table*}

We integrated 12 satellites whose centre of mass have initial conditions given in either chaotic or regular regions of the phase--space of the smooth host potential.  Since the accesible regions of the phase--space are too extense  to make a comprehensive dynamical study (i.e. they are six dimensional and unbounded), we decided to focus on the typical orbits that give birth to minor mergers.  This is done by imposing a condition on the initial circularity $\eta$ of the orbits.\footnote{We define this parameter as the ratio between the specific angular momentum of the orbit under consideration and that of a circular orbit (which is solution of a spherical NFW  potential with the same virial mass and concentration of the halo studied here) with the same specific energy.}  So, for all the orbits we fixed $\eta=0.54$ which is the most probable value obtained in the cosmological simulations of~\citet{2011MNRAS.412...49W}.
As we are interested in probing different galactocentric scales, we launch the orbits from apocentres at three typical distances: $30,~50$ and $80~\rm{kpc}$. Working with spherical coordinates $r$ (radius), $\theta$ (polar angle), $\phi$ (azimutal angle) and velocities ${\rm{v}}^{\alpha}$ ($\alpha=r,\theta,\phi$), this condition translates into:
\begin{align}
  \label{ic_2}
        {\rm{v}}^r_0 & = 0, \nonumber \\
        r_0 & = 30,~50,~80~\rm{kpc}, 
\end{align}
where the null subscript denotes an initial value.
These apocentric distances are inside the present host virial radius so they are different from the true apocentres at infall. We 
  interpret them as effective values compatible with the energy and angular momentum of the corresponding orbit~\citep{2011MNRAS.412...49W}.

For simplicity, we set all the initial velocity vectors to be parallel to the z=0 plane, so that
\be
\label{ic_2b}
{\rm{v}}^{\theta}_0 = 0.
\ee
Finally, considering the definition of circularity $\eta$ and Eqs.~\ref{ic_2}-\ref{ic_2b}, we arrive to an equation for the
equivalent circular radius $r_c$ together with an expression for the initial (azimutal) velocity:
\begin{align}
  0&=\frac{1}{2}\left[ 1-\frac{r_c^2}{r_0^2}\eta^2\right] \frac{f(r_c/r_s)}{r_c}-\frac{\ln(1+r_c/r_s)}{r_c} +\frac{\ln(1+r_0/r_s)}{r_0}, \nonumber \\
  {\rm{v}}^{\phi}_0& = \eta \frac{\sqrt{G r_c M_{200}}}{r_0} \sqrt{\frac{f(r_c/r_s)}{f(c)}}.
\end{align}
We have thus obtained three surfaces of initial conditions with each one being a sphere in configuration space with radius equal to $r_0$ and parametrized by the two angular variables, $\theta_0$ and $\phi_0$.

Now it is possible to study the phase--space structure of these surfaces through a classical chaos indicator.  Fig.~\ref{fig:LI_progenitors_rApo-all} shows colour coded the largest Lyapunov characteristic exponent $L$ at finite time \citep{1980Mecc...15....9B, 2010LNP...790...63S} in the plane ($\phi_0, \cos \theta_0$). 
The higher the value of this indicator, the more chaotic the orbit is.  These orbits and their associated variational equations (necessary to calculate $L$) were integrated for 1000 Gyr with the eight order symplectic integrator presented in \citet{2000CoPhC.130..176S}.  The time step chosen for the integrations was $\Delta t=0.1$~Myr.  We repeated the analysis using the smaller alignment index \citep{2001hell.confE...7S, 2003PThPS.150..439S, 2004JPhA...37.6269S}, which is an alternative way of measuring the chaoticity of a system.  Results obtained with both estimators are consistent.
Chaotic orbits amount to 49\%, 44\% and 32\%, respectively for each panel.
  For each apocentric distance, we selected two of the most chaotic orbits and two of the least chaotic (i.e. most regular) orbits.
Table~\ref{table_ics} shows the values of these initial conditions for the centres of the satellites.  The table also shows the radius of the equivalent circular orbits, the value of the Lyapunov exponent and the classification that we infer from it.  Only for the chaotic cases we include the Lyapunov time $T_{\rm{L}}$, which is an estimation of the time it takes for two infinitely close orbits to diverge from each other by a factor $e$. In other words, it is inversely proportional to the degree of chaos of the orbit. Besides, we give the value of the smallest pericentric distance
$r_{\mathrm{min}}$, the mean pericentric distance $\langle r_{\mathrm{p}} \rangle$ and the corresponding standard deviation.

\subsection{Structure of the progenitors}
\label{sec:init_cond_struc_sat}

The initial conditions for the $N=10^6$ dark matter particles of the satellite were drawn from a Plummer distribution~\citep{1911MNRAS..71..460P}, whose mass inside a given radius $r$ is given by:
\begin{equation}
  M_{\rm{\textsc{dm}}}(r) = \frac{M_{\rm{\textsc{dm}}}}{\left[1+ (b_{\rm{\textsc{dm}}}/r)^2 \right]^{3/2}}.  \nonumber
\end{equation}
We assumed the following values for the total  mass $M_{\rm{\textsc{dm}}}$ and the length scale $b_{\rm{\textsc{dm}}}$:
\begin{align}
M_{\rm{\textsc{dm}}} & = 1.6\times10^8~\rm{M}_\odot, \nonumber \\
b_{\rm{\textsc{dm}}} & = 0.62~\rm{kpc}, \nonumber
\end{align}
which are consistent with the Sculptor galaxy and were fixed by taking into account constraints on the dark matter mass at two different radii provided by \citet{2010MNRAS.406.1220W} and \citet{2013MNRAS.433.3173B}:
\begin{align}
  M_{\rm{\textsc{dm}}}(r_{1/2}=0.375~\rm{kpc}) & = 2.25\times10^7~\rm{M}_\odot, \nonumber\\
  M_{\rm{\textsc{dm}}}(1~\rm{kpc}) & = 10^8~\rm{M}_\odot, \nonumber
\end{align}
where $r_{1/2}$ is the half-light radius.

Although the evolution of the barionic component is not considered in the $N$-body simulations, the stellar matter can be tracked through an association with the most tightly bound material in the dark matter halo. Specifically, each dark matter particle is labelled with a given value of stellar mass applying the method of~\citet{2005ApJ...635..931B}, as follows.  We assume that the stellar component is initially distributed as an isotropic Plummer sphere of mass $M_{\star}$ and length scale $b_{\star}$ and that it does not contribute to the gravitational potential.  In this way, denoting the Plummer density and potential as $\rho_{\star}(r)$ and $\Phi_{\rm{DM}}(r)$, the stellar--to--dark mass ratio of a particle of relative specific energy $\epsilon=-(\Phi_{\rm{DM}}(r)+{\boldsymbol{\rm{v}}} \cdot {\boldsymbol{\rm{v}}}  /2)$ is given by 
\begin{equation}
  \label{Upsilon}
  \Upsilon(\epsilon)= f_{\star}(\epsilon)/f_{\rm{DM}}(\epsilon),  \nonumber
\end{equation}
where
\begin{align}
  f_{\rm{DM}}(\epsilon)& = \frac{24\sqrt{2} b_{\rm{DM}}^2}{7 \pi^3 G^5 M_{\rm{DM}}^4} \epsilon^{7/2}, \nonumber\\ 
  f_{\star}(\epsilon)&=
  \frac{1}{\sqrt{8}\pi^2}
  \int_0^\epsilon \frac{d^2\rho}{d\Psi^2}\frac{d\Psi}{\sqrt{\epsilon-\Psi}}, \nonumber 
\end{align}
are respectively the dark matter and stellar phase--space distribution densities
(rewritten as functions of $\epsilon$), with $\rho\equiv\rho_\star(r(\Psi))$, $\Psi\equiv \Psi_{\rm{DM}}=-\Phi_{\rm{DM}}$ and
\begin{equation}
  \label{satellite_dm_dens}
  \rho_{\star}(r) = \frac{3 M_{\star}}{4\pi b_{\star}^3} \frac{1}{\left[1+ (r/b_{\star})^2\right]^{5/2}}. \nonumber 
\end{equation}

\citet{2007ApJ...663..948G} showed that for any Plummer sphere that models matter in which mass follows light, the observed 2D projected half--light radius, $r_{\rm{2D}}$, is exactly the scale parameter $b_{\star}$.  Moreover, according to~\citet{2015ApJ...801...98S} or Table B1 of~\citet{2010MNRAS.406.1220W}, the 3D and 2D Plummer's half--light radii are related to each other by
\be
r_{1/2}=(2^{2/3}-1)^{-1/2}~r_{\rm{2D}} \nonumber
\ee
 so that 
\begin{equation}
  b_{\star}=r_{\rm{2D}}= (2^{2/3}-1)^{1/2} r_{1/2}\approx 0.766\times0.375\sim 0.29~\rm{kpc}, \nonumber
\end{equation}
which is close to the value used by~\citet{2013MNRAS.433.3173B}.  Regarding the stellar mass, we adopted $M_\star=10^6~\rm{M}_\odot$ as in~\citet{2013MNRAS.433.3173B}.


\subsection{$N$-body parameters}

The softening for the $N$-body simulations was chosen by performing the following three--step procedure. First of all, we applied the method used in~\citet{2008MNRAS.391.1806V}, obtaining a theoretical optimal softening parameter of $\tau \sim 16$ pc for our situation of $N=10^6$ dark matter particles. Afterwards, we checked that our realization of the Plummer sphere, satisfies the scalar virial theorem, obtaining $2T/|V|=1.00311$ for $\tau=16$ pc and $2T/|V|=0.99986$ for $\tau=0$ pc, where $T$ and $V$ are respectively the kinetic energy and the
virial\footnote{
  The virial for a gravitational system of $N$ particles with softening $\tau$ is equal to 
  \begin{equation}
    V~=~ -G \sum_{i=1}^{N-1}\sum_{j=i+1}^N 
    \frac{m_i m_j}{(r_{ij}^2+\tau^2)^{1/2}}
    +\tau^2 G \sum_{i=1}^{N-1}\sum_{j=i+1}^N 
    \frac{m_i m_j}{(r_{ij}^2+\tau^2)^{3/2}} \nonumber
  \end{equation}  
  where $r_{ij}\equiv|\boldsymbol{r}_i-\boldsymbol{r}_j|$, and $\boldsymbol{r}_i$ and $m_i$ are respectively the position and mass of the i$^{th}$--particle. Note that for a null softening, the virial is equal to the potential energy.
}
of the system.  At last, we verified numerically the stability during a Hubble time of our isolated progenitor.  We set the maximum time step to be $\Delta t _{\rm{max}}=1$~Myr and, following the standard criterion of \textsc{gadget-2}~\citep[Appendix A2 of][]{2008MNRAS.391.1806V}, we obtained an effective time step  of  $\Delta t _{\rm{eff}}=0.4$ Myr during the whole integration.

\section{Analysis}
\label{section:analysis}

The aim of our analysis is to determine how the presence of chaos affects our ability to detect tidal streams using data from surveys such as Gaia~\citep{deBruijne2012,2018JCAP...07..061B} or the future Large Synoptic Survey Telescope~\citep{2009arXiv0912.0201L}.  In order to do that, we need to specify both stream membership (i.e. whether stars in our simulations belong to the progenitor or form part of a stream) and a detectability criterion.

\subsection{Characterization of streams: membership and detectability}
\label{sec:char}

For each simulated satellite, we will define its associated stream in terms of the relative distance between the stellar particles and the centre of mass of the satellite in configuration space.  The centre of the satellite at time $t$ is defined as the instantaneous centre of mass of the five particles that, at the begining of the simulation, are the most bounded. According to the definitions given in the previous section, the degree of boundedness of each particle at the beginning of the simulation is computed using the Plummer potential.  At every time $t$, we define the stream as the set of particles that for any previous time $t'<t$ departed from the centre of the satellite in more than 5 kpc. The progenitor at time $t$ is defined as the complement of the stream; that is, as the set of particles that have always (for all $t' < t$) been inside a sphere of radius 5 kpc. We selected this value because it is large enough in order to include all the initial stellar mass of the progenitor and small enough in order to follow stellar streams as soon as they are formed.

\begin{figure}
        \centering      
\includegraphics[width=0.49\textwidth]{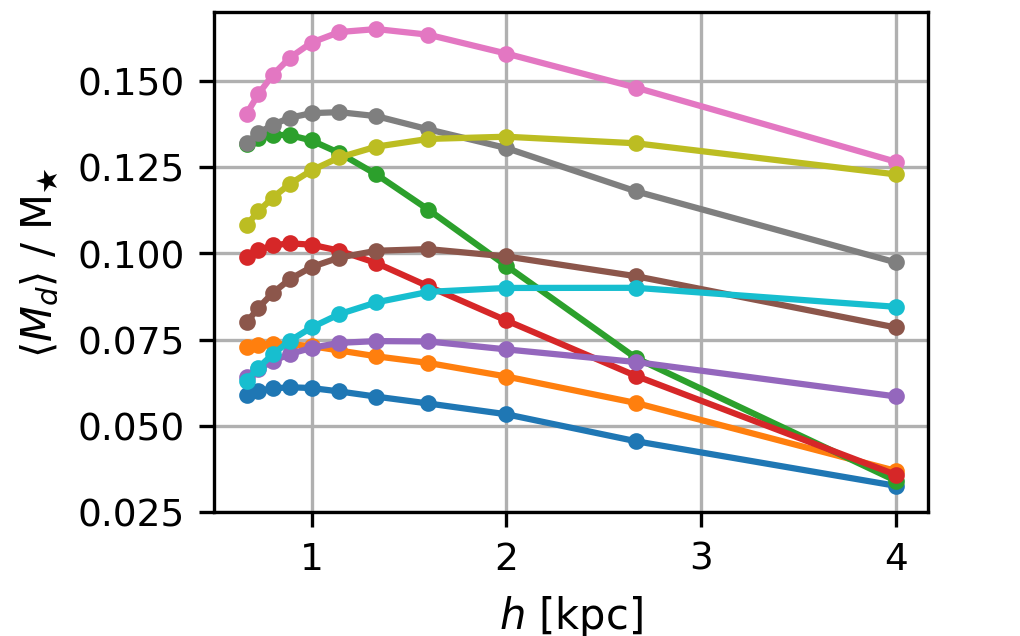}
        \caption{Convergence test of the detectability criterion.  Left: Dependence of the time average of the fraction of detectable stream mass, $\langle M_d \rangle$,   with respect to the initial stellar mass of the satellite, $M_{\star}$, as a function of the resolution, for satellites \#1 to \#10.}
        \label{fig:convergence}
\end{figure}
Next we define a detectability criterion based on configuration space data.  From an observational perspective we would observe the stellar mass distribution in configuration space of the total stellar halo, $\rho_{\mathrm{total}}$, and would compare it with a model of the background stellar halo distribution, $\rho_{\mathrm{halo}}$, trying to identify significant overdensities. We clasify a stream particle as detectable if
\be
\label{dn}
d \equiv \frac{\Delta \rho}{n~E(\Delta \rho)} > 1, 
\ee
where
\be
\label{delta_rho}
\Delta \rho = \rho_{\mathrm{total}} - \rho_{\mathrm{halo}}
\ee             
is the halo overdensity (i.e. stream density), $E(\Delta \rho)$ is an estimation of the error in the determination of $\Delta \rho$ and $n$ is the statistical significance of the detection.  These formulae are equivalent to saying that a stellar particle is detectable whenever the signal-to-noise ratio of the overdensity is larger than the desired detection significance.

The overdensity was estimated from the N-body simulations on a grid of size $h$ with a triangular shaped cloud (TSC) scheme~\citep{1988csup.book.....H}.  The selection mechanism of $h$ will be explained ahead in this section.  According to Eqs.~\ref{dn} and \ref{delta_rho}, in order to calculate $E(\Delta \rho)$, we need to take into account separate contributions from the estimations of the density of both the background stellar halo of the MW and the total stellar halo (which includes the stream particles):
\be
E(\Delta\rho) = \left[E^2(\rho_{\mathrm{total}}) + E^2(\rho_{\mathrm{halo}})\right]^{1/2}.
\label{two_errors}
\ee
To estimate the component $E^2(\rho_{\rm{halo}})$ associated
to the stellar halo,
we model the halo density with a powerlaw distribution:
\be
\rho_{\mathrm{halo}}(r) = \frac{A}{(r/R_\odot)^u},
\ee
where $A$ and $u$ are free parameters and $R_{\odot}$ is the Sun's
galactocentric distance, for which we take a standard mean
value $R_{\odot}=8~ \rm{kpc}$.
We assume that the two free parameters in this definition are independent of the
velocity and adopt mean values from~\citet{Hernitschek_2017}:
$(A,u)=(10^5~\rm{M}_\odot {\mathrm{kpc}}^{-3},3.5)$,
which were obtained from the distribution (renormalized) of RR Lyrae.  The associated variance can then be computed taking into account standard error propagation:
\begin{multline}
E^2(\rho_{\mathrm{halo}}) = \\
=  \left[\frac{\partial \rho_{\mathrm{halo}}}{\partial A}\right]^2 E^2(A) + \left[ \frac{\partial \rho_{\mathrm{halo}}}{\partial u}\right]^2 E^2(u) + \left[ \frac{\partial \rho_{\mathrm{halo}}}{\partial R_\odot}\right]^2 E^2(R_\odot) =  \\
=  \rho^2_{\mathrm{halo}} \left[ \frac{E^2(A)}{A^2} + \ln^2(r/R_\odot)E^2(u)  + \frac{u^2}{R_\odot^2}E^2(R_\odot) \right],
\label{drho}
\end{multline}
where $E(A)$, $E(u)$ and  $E(R_\odot)$ are the standard deviations of the fitted model parameters.  We will present our main results assuming hypothetical $1\%$ errors and will analyse the impact of larger errors in Sec.~\ref{sec:impact_chaos}.
According to Eq.~(\ref{drho}), $1\%$ errors in the parameters correspond to a total error in the density of 
  0.5--2.5\%, for distances between 10 and 80 kpc, respectively.

In order to model the error $E(\rho_{\rm{total}})$ that corresponds to the observed density, we consider Poissonian noise, while approximating the mass $m_\star$ of every star to be equal to one solar mass. As the number of stars in a given cubic cell of volume $h^3$ is
\be
N(\textbf{r})=\frac{h^3}{m_\star} \rho_{\mathrm{total}}(\textbf{r}),
\ee
and its shot noise is $E(N)=\sqrt{N}$, we have that the squared shot noise of the total density is
\be
E^2(\rho_{\mathrm{total}}) =\left[\frac{m_\star}{h^3} E(N)\right]^2 =\frac{m_\star}{h^3}\rho_{\mathrm{total}}.
\label{error_total}
\ee
The total density in this equation corresponds to the observed density. 
Since we are not working with a full mock halo catalogue, we estimated the total halo density with the addition of both the theoretical density of the stellar halo of the MW and the density of the mock stream:
\be
\rho_{\mathrm{total}}=\rho_{\mathrm{halo}} + \Delta \rho.
\ee
Once we obtained all the necessary components of the error and densities on the grid, we calculated the
detectability criterion defined in Eq.~\ref{dn} on the grid and interpolated back to the particles of the
simulation afterwards with the same TSC scheme employed for the density estimation.
Note that a similar criterion was applied to SDSS-Gaia Catalogue data by~\citet{2018MNRAS.475.1537M}
in the velocity space.

\begin{figure*}
  \centering
  \includegraphics[width=1.0\textwidth]{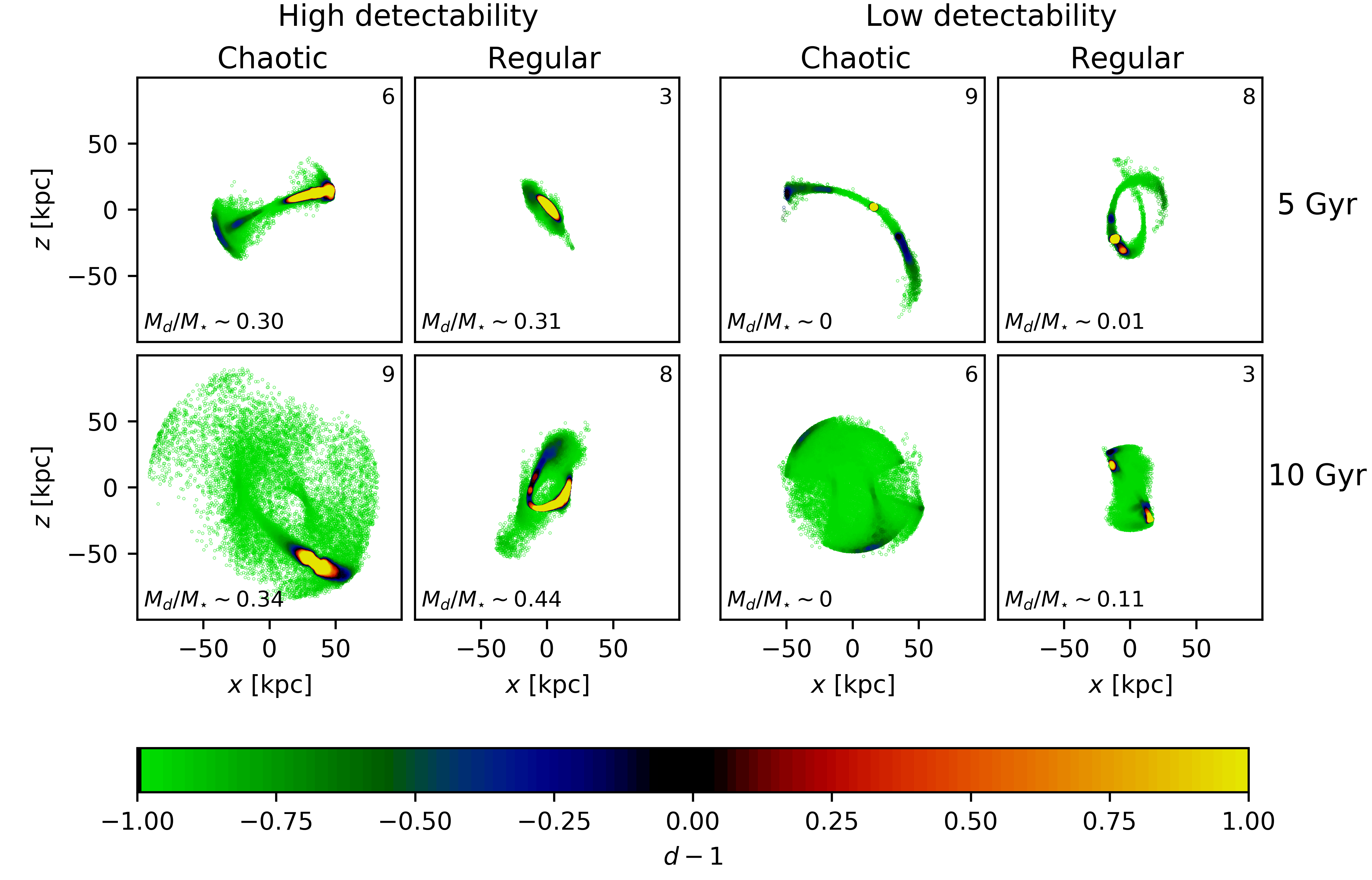}
  \caption{Examples of streams at times 5 Gyr (top) and 10 Gyr (bottom) with high (left) and low (right) detectable stream mass $M_d$. The projection shown is the $(x,z)$, colour coded with $d-1$. According to Eq.~\ref{dn}, the detectable stars are coloured red or yellow while  the undetectable stars are coloured blue or green. The satellite number is given in the top--right corner of each panel, and $M_d$ is given at the bottom of each panel. Note that among the yellow particles, the ones that belong to the progenitor do not contribute to $M_d$. We also classify the orbits of the progenitors as regular or chaotic.}
  \label{fig:figure_02_vx_vz}
\end{figure*}

For the significance parameter, we chose a value $n=4$, which implies a confidence in the detection of more than 99.99\%.  The only parameter that remains to be fixed is the grid spacing $h$.  While the error associated to the halo is independent of $h$, the stream density and the shot noise component of the error do depend on it (both of them grow when increasing the resolution). This implies that either the shot noise dominates (for small $h$ values) or the overdensity is very small (for large $h$).  So the detectability parameter decreases towards very high or very low resolutions.  In order to quantify these effects, we performed a convergence test whose result is summarized in Fig.~\ref{fig:convergence}. The points correspond to the time average of the fraction of stars that are detectable as a function of the spatial resolution as obtained from the simulations \#1 to \#10.  The values decrease towards very high or very low resolutions, having their maximum at $h\sim 1$ kpc, which is the value we use for the rest of our analysis.

  The detectability criterion we have defined corresponds to
  an idealized observation. Some of the characteristics
  of a true detection that are not considered by our model
  are: the use of local background volumes, use of generalized
  spaces with dimension higher than three,
  binning determined by the uncertainties
  of the observed quantities, samples restricted
  to  luminous evolved stars and  mass function dependence.

\section{Results}
\label{section:results}

We now summarize the results that we obtained from our simulations on the evolution of the streams an the impact of chaos on our ability to detect them.

\subsection{Visual appearance of the simulated streams}

Fig.~\ref{fig:figure_02_vx_vz} shows examples of four specific
streams (from satellites \#3, \#6, \#8 and \#9) at 5 and 10 Gyr,
projected in the plane $(x,z)$ and colour coded according to $d-1$.
Our algorithm classifies yellow and red particles as detectable, and blue and green ones as non detectable.
The black colour stands for the threshold situation.
In order to sample the particles in agreement with their stellar content, 
we have classified the particles into five stellar mass ranges in $M_\odot$ units: 0.4--1.5, 1.5--5, 5--10, 10--50, 50--120,
and plotted them using one every 20, 10, 5, 2, 1  particles, respectively. We have not plotted particles with
less than 0.4 $M_\odot$.
The top labels characterize whether the stream as a whole has high or low detectability, according to its total detectable stream
stellar mass $M_d$, which is written at the bottom of each panel.  It is worth noticing that satellites \#8 and \#9 at 5 Gyr have most of their detectable
stars concentrated in their respective progenitors so that these stars do not contribute to $M_d$ at all.

A priori, one would expect that the stars that escape from a progenitor
whose trajectory is in a chaotic region of phase--space will quickly diffuse
and not form a detectable stream.
However, we find very well defined streams in both chaotic and regular cases.
In fact, we can see that there are snapshots of chaotic and regular streams that look like similar to each other, with
regards to the detectability distribution in the sky.
This suggests that, if there is any effect of chaos on the detectability, it must be subtle.

\subsection{Three stages in the life of the streams}
\label{sec:stages_life}

\begin{figure*}
\centering
\includegraphics[width=\textwidth]{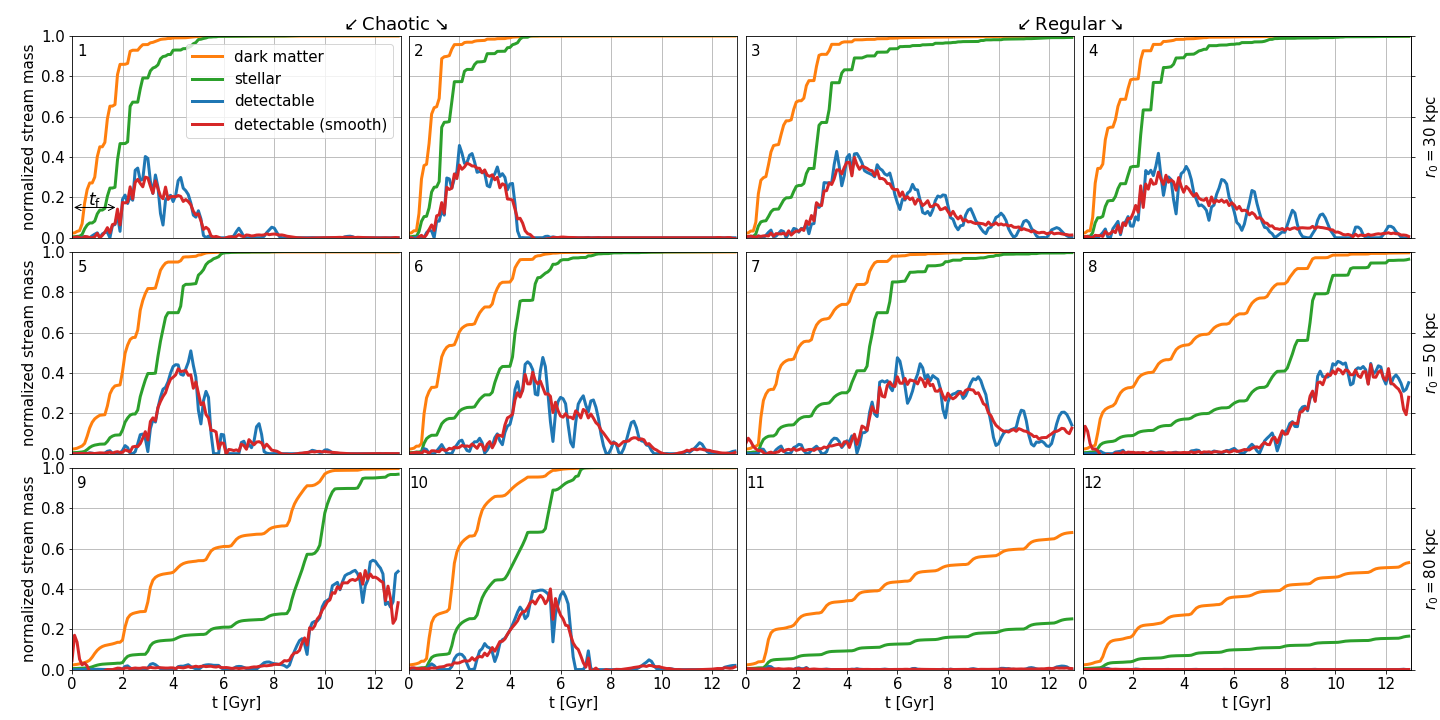}
\caption{Temporal evolution of stream masses: total dark matter (orange), total stellar (green),  detectable (blue) and filtered detectable (red), all of them normalized by the 
initial mass of the progenitor ($M_{\rm{DM}}$ or $M_\star$). The satellite number is placed in the top--left corner of each panel. The two columns to the left and the two to the right correspond to chaotic and regular orbits of the progenitors respectively. Each row corresponds to different apocentric radii of the initial conditions (smaller apocentres above).}
\label{fig:masses_of_t}
\end{figure*}

In order to better understand the results shown in Fig.~\ref{fig:figure_02_vx_vz} we analysed integrated quantities as a function of time.  The panels in Fig.~\ref{fig:masses_of_t} show the outcome of such analysis for each satellite separately.  The identification numbers shown in the upper-left corner of each panel correspond to the same IDs shown in Table~\ref{table_ics}.  The orange and green curves are the total dark matter and stellar masses of the stream, while the blue curves are the stellar mass of the stream that can be detected by our algorithm (i.e. $M_d$). All the curves are normalized with the corresponding initial mass of the progenitor ($M_{\star}$ or $M_{DM}$).
The strong oscillations that we find in the blue curves can be analysed considering that when the particles detach from the progenitor due to the
  host tidal force, their dynamics becomes approximately that of an ensemble of test particles in the halo of the host galaxy.  Thus, when a portion of stream is approaching
  its pericentre, it is expanded in coordinate space and when approaching its apocentre it is compressed.
Therefore, the stream density, and consequently its detectability, present oscillations whose period correspond to the radial orbital period of the satellite. As we are not interested in this phenomenon, but on the overall properties of the detectability of the streams, we filtered these oscillations in Fourier space using the following empirical kernel:
\begin{align}
W(\nu,c,s)=& w(\nu,c,s) w(\nu,-c,s), \nonumber \\
w(\nu, c, s) =& 1 - \frac{1}{1+\left[ (\nu-c)/s \right]^{10}},
\end{align}
where $\nu$ is the coordinate in Fourier space (i.e. frequency) and we assume $(c,s)=(1.5,1)~\mathrm{Gyr}^{-1}$. These values enable us to reduce the effects associated to the radial orbital frequency of the satellites and at the same time, maintain the main features of the detectability curves.  The resulting curves are shown in red in Fig.~\ref{fig:masses_of_t}.

Except for satellites $\#11$ and $\#12$, whose orbits are too far away from the host galaxy to form detectable streams, all the curves have similar qualitative behaviour.    It is possible to identify three distinct phases in the evolution
of the detectable part of the stream:
\begin{itemize}
\item \textit{Quiet formation:}  since the streams are formed by material that comes from the instantaneous outer layers of the progenitor and the stellar component is more concentrated than the dark matter one, most of the mass loss in the early phase belongs to the dark matter component of the satellites.  The release of stars is slow and thus, the stellar streams are not detectable.
\item \textit{Violent formation:}  when about 80\% of the total dark matter mass of the progenitor is lost, the remaining core still contains the 60\%
  of the initial stellar stellar mass of the progenitor ($M_\star$).
  So at this point, the mass of the detectable portion of the stream
  increases abruptly up to a fixed value. This process is further enhanced by the fact
  that the potential well decreases while the particles are tidally
  removed from the progenitor.
  
\item \textit{Diffusion\footnote{We use the term \textit{diffusion} to mean a process of spreading out or divergence of deterministic orbits with nearby initial conditions.
    Similar approaches have been used in the study of chaotic diffusion in Hamiltonian systems by~\cite{2018arXiv181205430G,2014PhyD..266...49C,2006AJ....132.2114C,2005A&A...439..375C}.}:}
  once the progenitor is completely destroyed, the flow of mass into the stream ceases. Diffusion in phase-space dominates, reducing the density of the stream, with a consequent decline of the detectability curves.  Most of the stellar stream mass becomes undetectable by the end of this last stage.
\end{itemize}
Thus, the detectable mass of a stream depends on both the rate of expulsion of stars from the satellite and the diffusion rate.  The larger the former and the smaller the latter during a given time interval, the more detectable mass we will have in the stream at the end of that interval.  It should be noticed that all the streams reach a detectability peak between 30\% and 40\% of $M_\star$.  A consequence of this is that only a lower bound of the  satellite's original stellar mass can be obtained when counting stream stars in surveys.

It would be interesting to asses how the scheme of three stages here presented is modified when the halo of the progenitor is cuspy instead of cored~\citep{2015MNRAS.449L..46E}.
\subsection{Impact of chaos in the life of the streams}
\label{sec:impact_chaos}

\begin{figure}
    \includegraphics[width=0.5\textwidth]{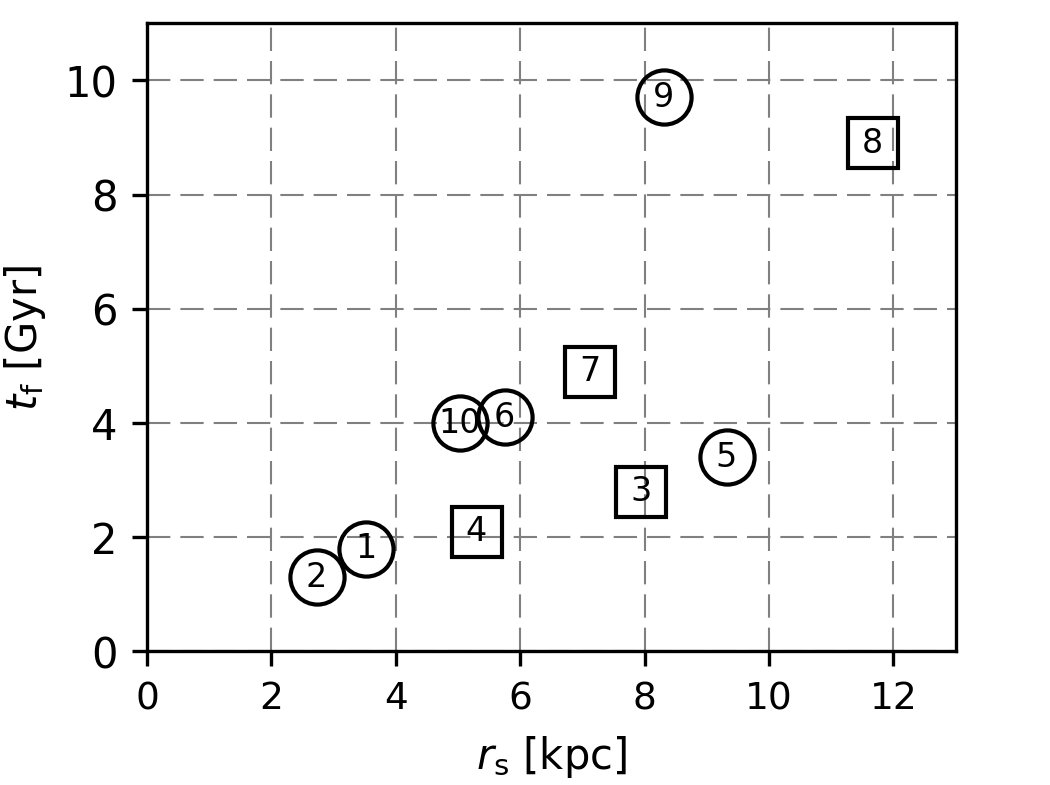}
    \caption{Formation time, $t_{\mathrm{f}}$, of stellar streams as a function of the smallest pericentric distance, $r_{\mathrm{s}}$, that takes place before $t_{\rm{f}}$. Each point in this figure corresponds to a different satellite, identified with a label.  Chaotic and regular satellites are marked respectively with circles and squares.  Since satellites \#11 and \#12 do not form detectable streams, we exclude them from the plot.}
    \label{fig:t_form_of_r}
\end{figure}

While the behaviour described above is the same for almost all the satellites, details of the transitions between different regimes depend on the orbital properties of the satellites (including their chaoticity).  For instance, the red curves in Fig.~\ref{fig:masses_of_t} show that the jump in detectable mass occurs at earlier times for satellites with smaller apocentric distances.  We quantify this by estimating the formation time of the streams $t_{\mathrm{f}}$ as the first instant in which the filtered detectability curves (red curves in Fig.~\ref{fig:masses_of_t}) reach half of their historical maxima.
  Fig.~\ref{fig:t_form_of_r} shows the dependence of this quantity with the smallest pericentric distance $r_{\mathrm{s}}$ that the satellite reached earlier than
  $t_{\mathrm{f}}$~\footnote{Note that although $r_{\mathrm{min}}$ (defined in Table~\ref{table_ics}) and $r_{\mathrm{s}}$ are
    similar quantities, they differ in two aspects:
  (i) the former  is computed using the symplectic orbit
    that corresponds to the initial condition for the progenitor while the latter is computed using
    the definition of the centre of mass of the progenitor given in Sec.~\ref{sec:char}; (ii) the former
     considers a Hubble time of integration while the latter considers only times up $t_{\mathrm{f}}$.}.
For this simulation suite, in which the initial internal structure of the progenitors is the same, the dominant effect that decides when the streams are formed is the typical pericentric distance of the orbits.
Progenitors in orbits with smaller pericentres feel stronger tidal forces, so that their associated Roche lobes shrink
faster, favoring the release of stars.
Consequently, orbits with smaller pericentres tend to form detectable streams faster than
orbits with greater pericentres. In Appendix~\ref{sec:appendix} we prove with high significance the existence of a weak correlation
between $L$ and $r_{\mathrm{min}}$ for a large sample of test orbits, which in combination with the $r_{\mathrm{s}}$--$t_f$ correlation implies 
the existence of a weak $L$--$t_f$ correlation (progenitors in chaotic orbits tend to disrupt faster than those in regular orbits).

\begin{figure}
    \includegraphics[width=0.5\textwidth]{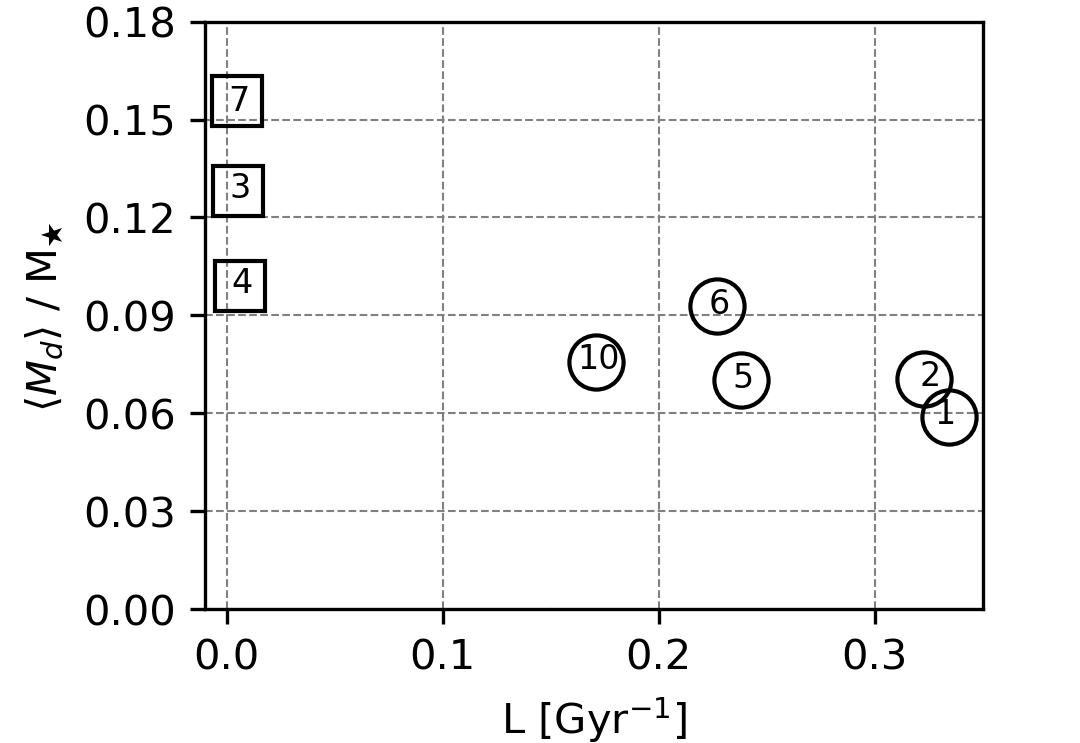}
\caption{Average detectability of stellar streams, $\langle M_d \rangle$, as a function of the Lyapunov exponent $L$. Each point in this figure corresponds to a different satellite, identified with a label. Chaotic and regular satellites are marked respectively with circles and squares.  The satellites \#8 and \#9 were not included in this figure because their formation time is so late that the regime of diffusion is not well developed.}
    \label{fig:avdet_of_l}
\end{figure}

Now we will study whether chaos has got any measurable contribution to the process of reduction of the stream observability.
The speed with which the detectability of the stream decreases after its historical maximum should depend on how fast the diffusion occurs in phase--space.  Since the stars that leave the progenitor behave essentially as test particles in the Milky Way halo, their posterior evolution depends on the properties of the galactic potential, which determines the diffusion rate.  Regions of the phase--space in which the orbits are regular or chaotic will be associated to diffusion with power law or exponential temporal dependencies respectively~\citep{1998MNRAS.301..960K}.  It is expected that streams that are originated in satellites that move in chaotic orbits will be more affected by this process and will reduce more rapidly the amount of detectable mass.  This will in turn also decrease the mean time interval in which the streams can be detected.  We confirm this reasoning by plotting the time average of the detectable mass $\langle M_d \rangle$ against the Lyapunov exponent $L$ of the trajectory of the progenitor (see Fig.~\ref{fig:avdet_of_l}).  The trend is that the larger the value of $L$ (i.e. the larger the degree of chaos in the surroundings of the progenitor), the smaller  $\langle M_d \rangle$, and so, the smaller the probability of detecting the stream.
  In order to have a quantitative estimate of the effect of chaos on the detectability, we have also computed the average of
  $\langle M_d \rangle$ over both the regular and the chaotic streams, finding values of 0.13 and 0.07, respectively. Thus,
  the presence of chaos reduces the amount of average detectable mass in a 50\% approximately.

\begin{table}
\caption{Correlation coefficients, $\mathcal{C}_x$ and $\mathcal{C}_v$, between
$L$ and $\langle M_d \rangle$ as a function of the relative error using
$\delta=E(A)/A=E(u)/u=E(R_\odot)/R_\odot=E(B)/B=E(\sigma)/\sigma$. The significance $\mathcal{L}_s$ is provided between braces.
}
\label{tab:error}  
\begin{tabular}{  l l l  } 
\hline
$\delta$ [\%]  & $\mathcal{C}_x$~$\{\mathcal{L}_s\}$ & $\mathcal{C}_v$~$\{\mathcal{L}_s\}$ \\
\hline
1 & -0.85~\{0.5\} &  -0.84~\{0.6\}\\
2 & -0.73~\{~3\}   &  -0.61~\{10\} \\
3& -0.62~\{~9\}   &  -0.37~\{36\}\\
4& -0.54~\{15\}  &  -0.24~\{56\}\\
5& -0.48~\{22\}  &  -0.09~\{85\}  \\    
\hline
\end{tabular}
\end{table}

In order to explore the correlation when increasing the value of the errors assumed for the parameters of the stellar halo distribution ($A$, $u$ and $R_\odot$), we computed $\langle M_d \rangle$ for a cubic grid of errors between 1\% and 5\% and found that  $\langle M_d \rangle$ decreases when increasing any of these three parameter errors. Table~\ref{tab:error} shows the Pearson correlation coefficient $\mathcal{C}_x$, between $L$ and $\langle M_d \rangle$, as a function of the relative error.
We also provide the level of significance $\mathcal{L}_s$, defined as the probability of rejecting the zero hypothesis (i.e. $\mathcal{C}_x=0$) when it is true.
Inspecting at this table it can be stated that for $\delta=1\%$ we can reject the zero correlation hypothesis with a certainty better than
99\%, while for $\delta=2\%$ this can be done at most with a certainty of 97\%.
When increasing the value of $\delta$, the significance of the correlation drops below 95\% which is our addopted threshold value.
For this reason, the test can not certify the existence of the correlation for values of the errors larger than 2\%.
Much significance could be gained by increasing the number of simulated streams.
Summing up, if the model of the stellar halo density in configuration space is known with a precision $\leq$ 2\%, then the probability of detecting a stellar stream is larger for regular orbits than for chaotic ones.

\subsection{Parallel results using the velocity space}
Milky Way surveys provide information not only in configuration space, but also in velocity space.
Thus, it may be worth asking whether our findings persist when performing the analysis in velocity space.
We have performed a parallel detectability analysis using velocity distributions as in~\citet{2018MNRAS.475.1537M}.
This time the halo density is modelled with a Gaussian distribution:
\be
\rho_\mathrm{halo}(v) = B\exp\left[-v^2/(2\sigma^2)\right]. 
\ee
Throughout the analysis, we assumed that the two free parameters in this definition are independent of the position in configuration space and addopting the values
$B=10^9/(2\pi\sigma^2)^{1.5}~\rm{M}_\odot(\mathrm{km/s})^{-3}$ and $\sigma=150~\mathrm{km/s}$, which is consistent with~\citet{2008MNRAS.391...14D} and
with~\citet{2018MNRAS.475.1537M}, respectively.

\begin{figure*}
\centering
\includegraphics[width=\textwidth]{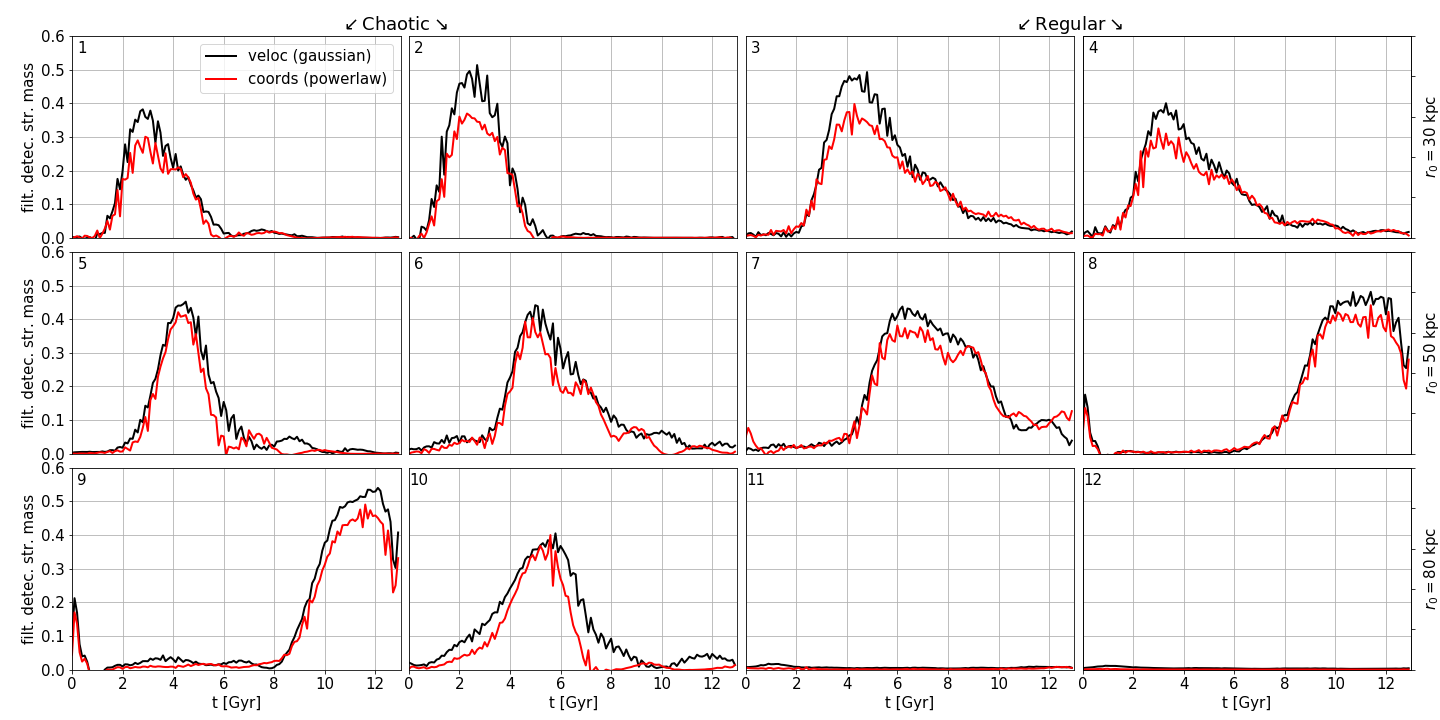}
\caption{Temporal evolution of filtered detectable stream mass using coordinate (red) and velocity (black) spaces, normalized with $M_\star$. The satellite number is placed in the top--left corner of each panel. The two columns to the left and the two to the right correspond to chaotic and regular orbits of the progenitors respectively. Each row corresponds to different apocentric radii of the initial conditions (smaller apocentres above).}
\label{fig:compare_criteria}         
\end{figure*}

Fig.~\ref{fig:compare_criteria} shows the time evolution of the filtered detectable mass of the coordinate approach (in red as in figure~\ref{fig:masses_of_t})
and the velocity approach (in black). We see that both behaviours are pointwise similar so that the three stages in the life of a stream presented
in Sec.~\ref{sec:stages_life} are recovered. Nevertheless, there is a slight variation in the maximum detectability peak attained, that in this
case is about 40\%--50\% $M_\star$.

\begin{figure}
    \includegraphics[width=0.5\textwidth]{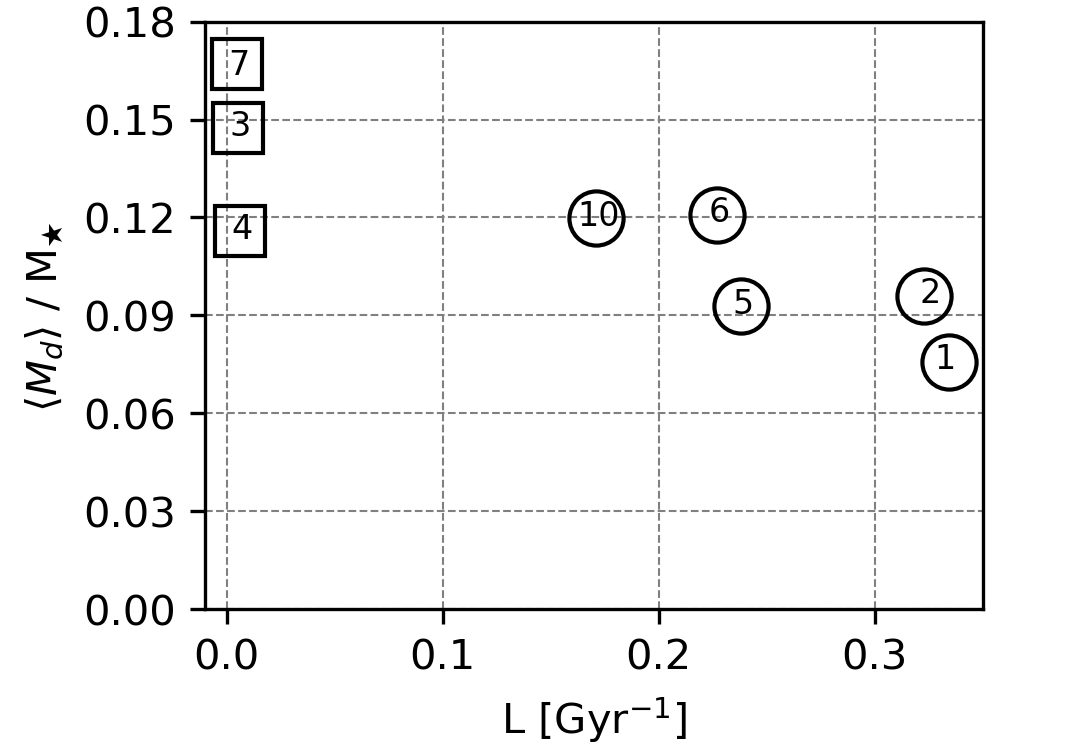}
\caption{Same as Fig.~\ref{fig:avdet_of_l} but using a detectability criterion in velocity space.}
    \label{fig:avdetvel_of_l}
\end{figure}

When comparing figures~\ref{fig:avdet_of_l} and~\ref{fig:avdetvel_of_l},  it can be noticed that the results on both spaces are closely similar.
The coefficient $\mathcal{C}_v$ in Table~\ref{tab:error} was computed analogously to $\mathcal{C}_x$ but in velocity space.
The results are compatible with those presented in Sec.~\ref{sec:impact_chaos}, where the significancy is better than 95\% as long as
$\delta\leq1\%$.
This is another example of the robustness of the detectability criterion presented in this work.

\section{Conclusions}
\label{section:conclusions}
The objective of this paper is to study theoretically the influence of chaos on
the detectability of Galactic stellar streams. For this purpose, we analysed a suite of 12 simulations of minor mergers. In all of them the host galaxy was modelled with a triaxial generalization of the NFW profile and the satellites were modelled with a Plummer sphere with parameters that resemble the Sculptor galaxy. The only difference between these experiments was the initial condition of the centre of mass of the progenitor. Half of these initial conditions were chosen to give birth to chaotic orbits while the rest were chosen so that their corresponding orbits were regular. The chaoticity had previously been determined by computing the largest Lyapunov characteristic exponent $L$ of each orbit. 

Each satellite was embedded in the host halo potential and was evolved with a self--gravitating $N$--body code (\textsc{Gadget-2}).  These tidal stream simulations were analysed with a detectability criterion that depends on the  stream stellar density and on a smooth model of the density of the stellar halo.  This criterion works independently with either configuration or velocity data, giving similar outcomes.

We found that the evolution of the detectability (detectable mass $M_d$) of the stellar stream goes through three distinct stages: quiet formation, violent formation and diffusion. During the initial phase the satellite looses most of the outer layers of its dark halo, but just a small proportion of its total stellar mass. The rate of stellar stream formation is low until the progenitor has lost about 80\% of its dark matter halo when this rate abruptly increases, setting in the beginning of the violent formation stage.  All the stellar streams reach now a detectability peak of magnitude between 30\% and 40\% of the original stellar mass of the satellite for the detectability criterion applied in configuration space. For the criterion in velocity space, this quantity ranges between 40\% and 50\%. These upper bounds imply that counting stream stars leads to a considerable underestimation of the initial mass of the progenitor.

Hamiltonian diffusion, which is always present and tends to wash out the stream structures, becomes dominant when the stellar injection stops due to the total disruption of the progenitor. This event settles the diffusion stage that is characterized by a decay of $M_d$. The fact that $M_d$ decays faster for a stream whose progenitor behaves chaotically than for those associated to regular orbits, is captured in the existence of a correlation between the Lyapunov exponent $L$ and the temporal average of the detectability $\langle M_d \rangle$. 

Through a significance analysis of the correlation, it was found that the previous result is valid
as long as the errors in the stellar halo model parameters are smaller than about $1\%$. In other words, with our criterion and model errors of 1\%, chaos reduces the detectability. Consequently, if a host galaxy has a significant proportion of chaotic orbits and if we are working with a model and star count data highly precise (e.g. $\lesssim 1\%$), the probability of detecting streams will be proportionally lower.

As the satellite destruction time
is correlated with the chaoticity through its dependence on pericentric distances,
the proportion of chaotic orbits should have also incidence in statistical studies
of the number of dSph satellites. For example, ensembles of progenitors in chaotic
trajectories will tend to disrupt and form streams faster than progenitors in regular ones.

\appendix

\section{Correlation between chaos and pericentric distance}
\label{sec:appendix}

\begin{figure*}
\centering
\includegraphics[width=0.8\textwidth]{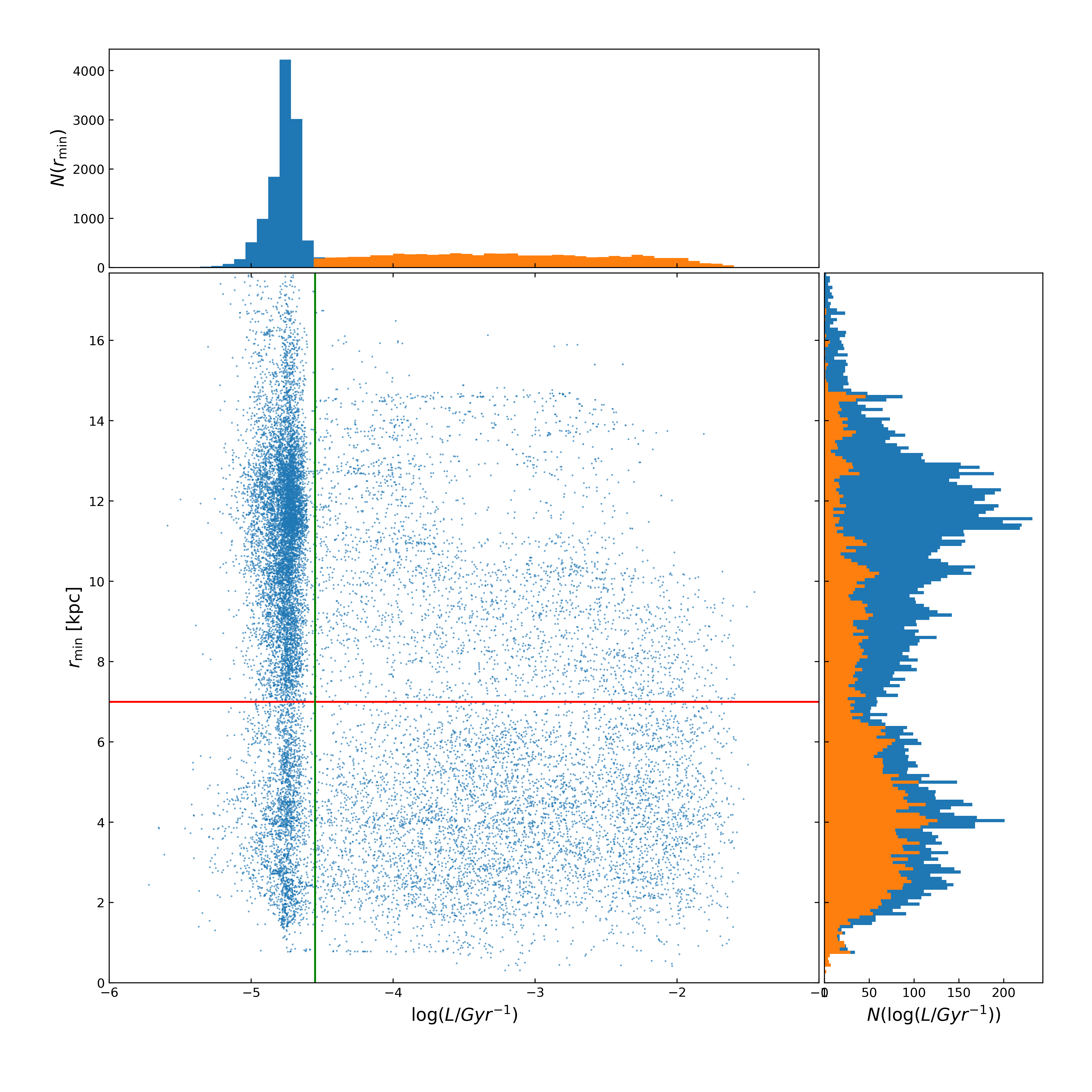}
\caption{$\log(L)$ versus $r_{\mathrm{min}}$ for 19810 test particle orbits whose initial
conditions belong to the bidimensional
surface with $r_{\textrm{apo}}=$50~kpc (middle panel of Fig.~\ref{fig:LI_progenitors_rApo-all}).
Marginal histograms of $L$ and $r_{\mathrm{min}}$ are shown in blue. The orange histograms correspond to the
chaotic orbits only (i.e. those at the right hand side of the vertical line).
}
\label{fig:lyap_rmin}
\end{figure*}
For 19810 test particle orbits that densely sample the orbits whose initial
conditions are displayed in the middle panel of Fig.~\ref{fig:LI_progenitors_rApo-all}
(i.e. for $r_{\mathrm{apo}}=50$ kpc) we have plotted their values of the largest
Lyapunov exponent at finite time $L$ versus their smallest pericentric distances attained
during a Hubble time $r_{\mathrm{min}}$, in the central panel of Fig.~\ref{fig:lyap_rmin}.
We also show the corresponding marginal distributions with blue histograms, noticing
a peaked clump of regular orbits for $\log(L)\lesssim -4.55$ ($L\lesssim 0.01$).
The  right hand side border of this clump can be used to divide the  $\log(L)-r_{\mathrm{min}}$ plane approximately
into chaotic and regular orbits, with the aid of a green line. 
Considering only the chaotic orbits, we have made
the corresponding marginal histograms in orange colour, from which it can be deduced that
chaotic orbits have a larger probability of reaching smaller pericentric distances than regular
ones.

In order to give futher support to this result, we have classified the points into
four sets according to the threshold values of
$\log(L)=-4.55$ and $r_{\mathrm{min}}=7$ kpc (horizontal red line).
We counted the number of points in each set (clockwise and starting from the top--left set):
8949, 2959, 5380 and 2522, respectively.
Moreover, we have computed the correlation coefficient between $L$ and $r_{\mathrm{min}}$, obtaining a value of $r=-0.34$.
    and the Student test gave a significance level of $\alpha=0\%$, implying that there is certainly a non zero 
    correlation between both quantities.

\section*{Acknowledgements}

CLL acknowledges support from STFC consolidated grant ST/L00075X/1 \& ST/P000541/1 and ERC grant ERC-StG-716532-PUNCA.
MM and DDC acknowledge support from CONICET (PIP0436), UNLP (G127)
and the program Erasmus Mundus EURICA. 
CLL thanks Shaun Cole for discussions.
MM thanks Amina Helmi, Shoko Jin, Tjitske Starkenburg and Nicol{\'a}s Maffione
for discussions.
We thank the referee for the constructive comments.




\bibliographystyle{mnras}
\interlinepenalty=10000
\bibliography{references} 


\bsp	
\label{lastpage}
\end{document}